# Effective connectivity signatures in major depressive disorder: fMRI study using a multi-site dataset

Peishan Dai[a*], Yun Shi[a], Tong Xiong[a], Xiaoyan Zhou[a], Shenghui Liao[a], Zhongchao Huang[b], Xiaoping Yi[c*], Bihong T. Chen[d].

[a]School of Computer Science and Engineering, Central South University, Changsha 410083, Hunan, P.R. China

[b]Department of Biomedical Engineering, School of Basic Medical Science, Central South University, Changsha 410013, Hunan, P.R. China

[c]Department of Radiology, Xiangya Hospital, Central South University, Changsha 410008, Hunan, P.R. China

[d]Department of Diagnostic Radiology, City of Hope National Medical Center, Duarte, CA, United States

***Corresponding authors:**

**Peishan Dai,** Ph. D, School of Computer Science and Engineering, Central South University, Changsha, Hunan 410083, P.R. China.

**Email:** daipeishan@csu.edu.cn

Telephone: 86-13787096016

**Xiaoping Yi**, Department of Radiology, Xiangya Hospital, Central South University, Changsha 410008, Hunan, P.R. China.

E-mail: yixiaoping@csu.edu.cn

Telephone: (011)86-731-84327488; Fax: (011)86-731-84327488

**Running title:** Effective connectivity for MDD

**Funding:** This research was funded by the National Natural Science Foundation of China (No. 82372039) and the Natural Science Foundation of Hunan Province of China (No. 2023JJ30695).

**Conflict of Interest:**

The authors declare no conflicts of interest.

**Abstract**

Diagnosis of major depressive disorder (MDD) primarily relies on the patient's self-reported symptoms and a clinical evaluation. Effective connectivity (EC) from resting-state functional magnetic resonance imaging (rs-fMRI) analysis can reflect the directionality of connections between brain regions, making it a candidate method to classify MDD. This study used Granger causality analysis to extract EC features from a large multi-site MDD dataset. The ComBat algorithm and multivariate linear regression were used to harmonize site difference and to remove age and sex covariates, respectively. Two-sample t-tests and model-based feature selection methods were used to screen for highly discriminative EC features for MDD, and LightGBM was used to classify MDD. In this large-scale multi-site rs-fMRI dataset, 97 EC features deemed highly discriminative for MDD were screened. In the nested five-fold cross-validation, the best classification model with the 97 EC features achieved accuracy, sensitivity, and specificity of 94.35%, 93.52%, and 95.25%, respectively. In another independent large dataset, which tested the generalization performance of the 97 EC features, the best classification models achieved 94.74%, 90.59%, and 96.75% for accuracy, sensitivity, and specificity, respectively. This work demonstrated that EC had a reasonable discriminative ability and supported the notion for using EC to potentially assist clinical diagnosis of MDD.

## 1. Introduction

Major depressive disorder (MDD) is a mental illness characterized by low mood and lack of interest in normally enjoyable activities[1]. Currently, the diagnosis of MDD is based on the patient's self-report and mental state examination, which do not necessarily reflect the patient's underlying brain activity[2][3][4][5]. Resting-state functional magnetic resonance imaging (rs-fMRI), a non-invasive technique for observing human brain activity, is increasingly used as a research tool to classify MDD, but its potential as a diagnostic tool has not been fully explored[6][7].

Functional connectivity (FC) from rs-fMRI has been used to classify MDD[8]. FC uses Pearson correlation to calculate the correlation between two brain regions, which represents how different brain areas communicate and work together to perform various cognitive and emotional processes. MDD patients have abnormal FC compared to healthy control (HC) groups[9][10], indicating that rs-fMRI data may be useful for classifying MDD. FC is also widely used in the classification of other neurological disorders[11][12][13]. However, the sample sizes have been relatively small in studies using FC for MDD classification. Thus, the generalizability of the classification model needs further testing. In addition, the classification performance of the FC was not high enough for clinical diagnosis of MDD [14][15][16][17]. Gallo et al. (2023) [68] extracted FC features from large-sample, cross-site datasets containing1249 patients with MDD and 1249 HC participants. They used Support Vector Machine (SVM) and Graph Convolutional Network (GCN) classifiers to achieve a mean classification accuracy of 61% for MDD versus the HC participants. FC is non-directional and only reflects the strength of correlation between two brain regions but not the direction of information transfer. Therefore, when applying FC to classify psychiatric disorders such as MDD, it may be challenging to achieve high classification performance.

Effective connectivity (EC) is another rs-fMRI-derived measure that represents the transfer of information from one brain region to another [18]. It is used to study the direction and strength of information transfer and how it contributes to the implementation of cognitive and emotional processes. EC can be obtained using techniques such as Granger causality analysis (GCA), which infers the causal relationship between time series data, such as neural activity recorded from different brain regions [19][20]. GCA-based EC has been widely used in fMRI studies to reveal causal relationships between brain regions [21][22][23]. Liu et al. found that the EC between the right ventrolateral prefrontal cortex and the left temporal-parietal junction in the ventral attention network is related to depression [24]. Other studies have found multiple EC features associated with depression [25][26][27]. Geng et al. obtained good classification results for MDD by using only the EC extracted from the default mode network (DMN), dorsal attention network (DAN), frontal-parietal network (FPN), and salience network (SN) on a small sample dataset [28].

In this study, we used GCA to extract EC from the large sample, cross-site MDD dataset "REST-meta-MDD Consortium". We selected features with significant discriminative ability, and used machine learning classifiers to classify subjects with MDD or being healthy subjects. We also performed data mining for important EC features that were associated with MDD.

## 2. Materials and Methods

*2.1 Participants*

For this study, we built and tested a classification model using the REST-Meta-MDD Consortium (http://rfmri.org/REST-meta-MDD), a cross-site large sample dataset of Chinese people, and extracted MDD-associated EC features. We then used the extracted MDD-associated EC features to test the generalization performance in the DecNef Project Brain Data Repository (https://bicr-resource.atr.jp/srpbsopen/), a cross-site large sample dataset of Japanese people.

*2.1.1 The REST-Meta-MDD Consortium dataset for extracting MDD-associated EC features*

The dataset used in this research was collected by the REST-Meta-MDD Consortium. This dataset consisted of 1300 MDD cases and 1128 HCs from 25 sites [51].

In order to ensure the quality of the data, we followed the steps shown in Figure 1 for data selection. We added the sample discarding step into the data screening process of Yan et al. [51] to exclude the samples with missing region-of-interest [ROI] signals. As a result, 832 MDD cases and 779 HCs from 17 sites were included in the final analysis for this study. Figure 1 presents the flow chart for case screening and exclusion. The demographic information of all participants is summarized in Supplementary Table S1.

*2.1.2 The DecNef Project Brain Data Repository dataset for testing the generalization performance of extracted MDD-associated EC features*

This data was obtained from the DecNef Project Brain Data Repository (https://bicr-resource.atr.jp/srpbsopen/) gathered by a consortium as part of the Japanese Strategic Research Program for the Promotion of Brain Science (SRPBS) and was supported by the Japanese Advanced Research and Development Programs for Medical Innovation (AMED). The demographic information for all participants is summarized in Supplementary Table S2.

*2.2 Data preprocessing*

The DPARSF software [52] was used to preprocess the data (http://www.rfmri.org/DPARSF). The following steps were included: (1) discarding the first ten volumes of the rs-fMRI data to eliminate the effects of magnetic field instability; (2) applying slice-time correction and realigning the image using a six-parameter (rigid body) linear transformation; (3) co-registering the T1-weighted images to the mean functional image using a six degrees-of-freedom linear transformation without re-sampling; (4) segmenting the T1-weighted images into gray matter, white matter, and cerebrospinal fluid; (5) normalizing all images to the Montreal Neurological Institute (MNI) space; (6) regressing head motion effects using the Friston 24-parameter model [53]; and (7) filtering all the time series data using a temporal bandpass filter (0.01~0.1Hz).

*2.3 EC extraction*

The rs-fMRI data were first preprocessed, and then the automated anatomical labeling (AAL) atlas was used to parcellate the entire brain into 116 regions [54]. The time-series datasets for each of the 116 brain regions were obtained by averaging the time series data for all the voxels within the brain region.

We used the REST-GCA software to perform EC calculations based on the signed path coefficient algorithm [20]. For REST-GCA, a Granger causality relationship model was used to calculate the EC between time series datasets, representing the activity of different brain regions. The positive or negative sign of the path coefficient indicated that the activity of one brain region could predict an increase or decrease in the activity of another brain region. This increase or decrease may indicate an excitatory or inhibitory effect of one brain region on another. The magnitude of the EC value represented the strength of directed connections between different brain regions.

The GCA model was used to calculate the EC between each pair of the 116 time-series datasets extracted using the AAL atlas. We obtained one EC matrix with the dimensions of 116*116, including two EC values for a pair of time series datasets: the EC from brain area x to brain area y, and the EC from brain area y to brain area x, corresponding to the upper right and lower left triangles of the matrix respectively.

*2.4 Site variance harmonization and biological covariate suppression*

To reduce the effect of site differences on classification performance, the ComBat algorithm (https://github.com/Warvito/neurocombat_sklearn) was used to harmonize site differences [55][56][57]. To reduce the effect of biological covariates on classification performance, multivariate linear regression was performed to remove age and sex covariates. Scikit-learn's "PowerTransformer" function was used to stabilize variance and minimize skewness in order to normalize the EC features (https://scikit-learn.org/stable/) [58][34].

Multivariate linear regression was used to reduce the influences of age and sex on the features [59]. The multivariate linear regression function was defined with the following formula:

$$\mathrm{Feature}_{(i,j)} = \beta^{o}_{(i,j)} + \beta^{age}_{i} * \mathrm{age}_{j} + \beta^{sex}_{i} * \mathrm{sex}_{j} \quad (1)$$

where $i \in \{1, \cdots, N\}$ was the feature index, $j \in \{1, \cdots, M\}$ was the participant index. $\beta^{o}_{(i,j)}$ was the feature value of the jth participant after removing the age and sex covariates. $age_j$ and $sex_j$ represented the age and sex of the jth participant, respectively. $\beta^{age}_{i}$ and $\beta^{sex}_{i}$ represented the age and sex regression coefficients of the i th model, respectively. The entire EC extraction procedure is shown in Figure 2.

*2.5 Feature selection and classification model*

Because the number of extracted features were much higher than the number of samples, feature selection was needed to reduce the dimensionality of the features. The feature selection and classification models used here were similar to those used in our previous study wherein we classified depression and extracted biomarkers using FC and network attributes [10]. We first used a two-sample t-test to select features with p-values <0.05, and to prevent information leakage, this step was only performed on the training set. This two-sample t-test can greatly compressed the feature dimensionality and eliminated a large number of invalid features [60][61]. Then, we used LightGBM [29] and SVM-RFECV implemented by Scikit-learn [34] to select important features. Both were common model-based feature selection methods [62][63], which achieved good results in classifying MDD using FC [10].

*2.6 Feature selection and classification method steps*

We integrated the procedures of feature selection and classification by incorporating a machine learning model into the feature selection procedure. We employed a nested five-fold cross-validation strategy during both the feature selection and classification procedures. The folds in the inner loop were used to tune the hyperparameters of the models.

The outer folds were used to extract the important features and evaluate the classification performance, respectively. The detailed steps of the method were as follows:

**Step 1: Feature selection.**

Because the dimensionality of the extracted features was very high, we needed to select the features that contribute greatly to the classification model. Dimensionality reduction methods such as PCA [64] were not used because we intended to identify which features contributed greatly when the model achieved the best classification performance, and then these features could be regarded as the EC features that were different between MDD cases and HCs. The steps of the feature selection procedure were as follows:

1. The data set was partitioned into a training set and a test set using an outer loop of a five-fold cross-validation strategy.

Steps 2 to 7 were repeated for each partitioned dataset.

2. A two-sample t-test with a p-value $<0.05$ was used to filter out the screened EC in the training set. The features in the test set were set to be the same as those in the training set.

3. In the training set (performing an inner loop of five-fold cross-validation), the best hyperparameters of the feature selection model (LinearSVM/LightGBM), using the features screened out by the two-sample t-test, were obtained using the hyperparameter optimization algorithm of Optuna [65].

4. In the training set, Imbalanced Learn's Synthetic Minority Oversampling Technique (SMOTE) [66][67] was used to maintain a balance in the sample proportion so that the ratio of MDDs to HCs was close to 1:1. The test set was not oversampled.

5. The oversampled training set was used to train the feature selection model (LinearSVM/LightGBM) with the best hyperparameters.

6. The test set was used to test the performance of the models. The test results were saved for comparison with the classification model retrained with important features later.

7. In the training set, SVM-RFECV was used to obtain an important feature subset for the LinearSVM model. The "feature_importance_" function was used to select important features for the LightGBM model, which was another important feature subset.

8. When the outer loop of five-fold cross-validation was complete, there were five important feature subsets for the LinearSVM and LightGBM models, respectively. The intersections of the five feature subsets were used as the final features.

Figure 3 shows the steps of the important feature subset extraction procedure.

**Step 2: Classification model training and testing.**

1. The data set was partitioned into a training set and a test set using an outer loop of a five-fold cross-validation strategy.

Steps 2 to 3 were repeated for each partitioned dataset.

2. In the training set (performing an inner loop of five-fold cross-validation), the best hyperparameters of the classification model (LinearSVM/LightGBM) using the final features selected by Procedure 1 (Feature selection) were obtained using the hyperparameter optimization algorithm of Optuna [65].

3. The test set was used to test different models and obtain testing results.

4. The model with the best performance was selected as the model used for MDD classification.

**Step 3: Contribution analysis of the selected features**

1. The classification model from step2 (Classification model training and testing) was selected.

2. In the outer loop of the five-fold cross-validation, Shapley Additive exPlanations (SHAP) was used to analyze the contribution of each feature to each classification model.

*2.7 Hyperparameter settings*

The feature selection and classification model involved many hyperparameters. For feature selection, we first performed a two-sample t-test on all features, retaining features with p-values below 0.05. Model-based feature selection methods were then used. The "sklearn.feature_selection.RFECV" function was used for feature selection, which was set to remove the 10% most unimportant features at each iteration; "scoring" was set to "balanced_accuracy;" and "cv" was set to "RepeatedStratifiedKFold (n_splits=5, n_repeats=1)." LightGBM used the "feature_importances_" function for feature selection, and the "importance_type" parameter was set to the default option "split," retaining features with "feature_importances_" values ≥1.

For the classification model, we used Optuna to perform a hyperparameter search for the hyperparameters of the classification model. We did five-fold cross-validation on the training set, using the accuracy of the five-fold cross-validation on the training set as the target for Optuna tuning optimization.

We set the Optuna tuning iterations to 100 ("n_trials" was set to 100) and limited the time to 10 minutes ("timeout" was set to 600 seconds) for LinearSVM and LigthGBM. (LinearSVM's "max_iter" was set to ten million.). The detailed search space for the hyperparameters of LinearSVM is shown in Supplementary Table S3, while the detailed search space for the hyperparameters of LightGBM is shown in Supplementary Table S5.

## 3. Results

### *3.1 Comparison of the classification performance across classification models*

We compared the classification performance of the LightGBM [29] model with that of commonly used machine learning and deep learning methods. To ensure comparability, all EC experiments used the 13340 EC in the EC matrix, and all FC experiments used the 6670 FC in the FC matrix. The LightGBM classification method was more optimal than the traditional machine learning methods such as Rbf-SVM, Random Forest, and XGBoost (Table 1) [30]. Our classification method was also better than the other deep learning classification methods such as GCN [31], BrainNetCNN [32], and Transformer [33]. Additionally, we observed that models trained with EC generally outperformed those trained with FC. This trend was evident across the experimental results. Table 1 presents the comparison of the classification performance of machine learning and deep learning models.

The comparison of the receiver operating characteristic (ROC) curves of the machine learning and deep learning models is shown in Figure 4.

### *3.2 Impact of feature selection on the classification performance*

In the feature selection procedure, we used both SVM-RFECV implemented by Scikit-learn [34] and LightGBM for feature selection, but SVM-RFECV selected fewer EC features. Therefore, we used LightGBM's "feature_importances_" function to select important features, which were taken to be features highly discriminative for MDD, for training the LinearSVM and LightGBM models. The comparison of the classification performance of LinearSVM and LightGBM classifiers before and after feature selection is shown in Table 2.

As noted in Table 2, after feature selection, the number of EC features decreased from 13340 to 97, greatly reducing the dimensionality of the features. After feature selection, both the LinearSVM and LightGBM classification models had significantly improved classification performance. The LightGBM model, which used feature-selected EC, obtained the best classification results, achieving an accuracy, sensitivity, and specificity of 94.35%, 93.52%, and 95.25%. The ROC curves for the two classifiers before and after feature selection is shown in Supplementary Figure S1.

*3.3 Generalizability of the highly discriminative EC features*

To test whether our extracted highly discriminative EC features had good generalizability performance, we used the dataset from DecNef Project Brain Data Repository (https://bicr-resource. atr.jp/srpbsopen/).

Table 3 presents the generalizability of highly discriminative effective connectivity features. As shown in Table 3, the 97 highly discriminative EC features we extracted showed excellent classification performance on the DecNef Project Brain Data Repository dataset, indicating reasonable generalizability. The ROC curve of migration experiments with highly discriminative EC features is shown in Supplementary Figure S2.

*3.4 Comparison of the results of the proposed method with other studies using FC or EC features for MDD classification*

Table 4 shows a comparison of recent research results on classifying MDD using FC or EC features. Our method achieved the best classification results by using the largest sample size. Also, most of the prior MDD classification studies have primarily used FC features. One study used EC for MDD classification (Geng et al. [28]), but the sample size was small and the classification performance of the study was lower than that of our method.

In our previous research (Table 4, Dai et al. [10]), we used the same dataset as this study to mine important FC and network attribute features, and obtained 68.90% classification performance in the five-fold cross-validation scenario. However, if we only used FC without network attribute features, we could only achieve 67.10% classification performance. We achieved a 94.35% classification accuracy using EC features in this study, which significantly improved the classification performance for MDD.

*3.5 Contribution analysis of important features*

We used SHAP to assign the contribution of each feature to the best classification model. As we used nested five-fold cross-validation to train and test the classification model, we used SHAP to assign the contribution of features to the model on each fold of the five-fold cross-validation (outer loop). Figure 5 shows the results of the feature contribution assignment for the first fold in the five-fold cross-validation. We also provided the feature contribution assignments of the other four folds in the five-fold cross-validation in Supplementary Figures S3-S6.

We obtained five results of the feature contribution assignment in the nested five-fold cross-validation procedure. In Table 5, we counted the EC features with ≥3 occurrences for the five results.

Based on the data presented in Table 5, a visualization of the brain was generated using BrainNet Viewer [41], as depicted in Figure 6.

We used LightGBM to extract 97 EC features that had excellent discrimination ability. Notably, EC features originating from the cerebellum made up the majority, particularly the 107th ROI-Cerebellum_10_L and 102nd ROI-Cerebellum_7b_R in AAL.

The details of our 97 highly discriminative EC features for MDD are provided in Supplementary Table S4.

## 4. Discussion

In this study, we extracted EC features from 116 brain regions across the entire brain, achieving high-precision classification of the MDD and HC groups in a cross-site, large-sample dataset. We have also mined EC features that were highly discriminative for MDD. The results of our classification model on the large-sample dataset indicated a new benchmark for classification performance in this line of research.

We used machine learning for feature selection because it helped us understand which brain connections were most important in MDD. This approach made it more reasonable to interpret our results and to understand how MDD affected the brain. We focused on effective connectivity (EC) instead of functional connectivity (FC) because EC could identify the direction of communication between brain regions while FC could not. This was important for understanding how information moved in the brain, especially in conditions like MDD. We used cross-validation in our analysis to assure our findings being stable and reliable, which minimized the impact of random factors during the evaluation indicators of the model. We tested the generalization performance of the EC features in another independent large dataset to validate the model performance.

Our study showed that EC from rs-fMRI had better performance compared to FC. Gallo et al. (2023) [68] used the same dataset as we did, extracting FC features for the classification of MDD and HC, achieving only 61.27% accuracy, 61.38% sensitivity, and 61.31% specificity, as shown in Table 4. Although MDD classification using EC was less common than using FC, we found EC achieving good classification results. Geng et al.[28] also showed that EC had a stronger MDD classification ability than FC albeit in a smaller sample (24 MDD cases and 24 HCs). Our research results support the notion that EC with directional information may contain more brain information than FC with only correlational information.

Achieving high classification performance on a large-scale, cross-site dataset of MDD cases and HCs has been the main focus of using rs-fMRI data to assist MDD clinical diagnosis. The studies using FC for MDD classification on smaller samples may face issues with generalizability, while FC studies using large-sample, cross-site data generally show lower classification performance. On the other hand, EC reflecting the brain's directional connections, may contain more information for identifying MDD, which may ameliorate the low performance issue with FC. Here, we achieved high classification performance with EC that far surpasses that of FC in two large-scale, cross-site datasets. Thus, our results supported EC being potentially useful to assist in the clinical diagnosis of patients with MDD, and a similar approach may also benefit the identification of disease subtypes and prediction of response to treatment.

There are multiple, distinct resting state networks, which include the default mode network (DMN), visual network (VN), sensorimotor network (SMN), attention network (AN), salience network (SN) and frontal-parietal network (FPN) [28][42][43]. Our research showed that the number of EC features across the networks was significantly higher than the number of EC features within the networks, which indicated that MDD had a stronger impact on inter-network EC than that on intra-network EC.

We found that MDD may alter EC between some brain regions, particularly in the cerebellar regions. In the EC features that we screened, most had cerebellar regions as the starting point. These were mainly concentrated in Cerebellum_7b_R and Cerebellum_10_L. The cerebellum is involved in motor function as well as higher-order functions, including cognition and emotion [44][45][46], fear memory [47], and emotional processing [48]. Abnormalities in cerebellar structure and FC have been found in several major psychiatric disorders, such as bipolar disorder, schizophrenia, generalized anxiety disorder, obsessive-compulsive disorder, and MDD [26][49][50]. The changes in EC found in our study of MDD patients further supported the understanding that changes in cerebellar function were closely related to MDD.

There were several limitations to this study. First, this study was limited as we focused on EC features for classification of MDD. Combining FC and other features may further improve classification performance. Second. the samples used in this study were from China, which may hinder the generalizability of the results as differences in populations may affect rs-fMRI data. In future studies, we will include rs-fMRI data from other populations as well. Third, this study only analyzed data on MDD and the classification models were not tested in other mental disorders.

In summary, we explored the alterations of brain EC in MDD based on a large cross-site rs-fMRI dataset. We found that EC had a better discriminative ability for MDD than FC. The MDD classification model we established had significantly better performance than other large cross-site MDD classification studies. In addition, we observed that discriminative EC features were mainly concentrated in the cerebellum, and most of them started from the cerebellum, which indicated that the information transfer from the cerebellum to other brain regions being closely related to MDD. This finding indicated a need to further investigate the relationship between EC of the cerebellum to cerebral regions in patients with MDD. Furthermore, our data showed that cross-network connections were significantly higher than intra-network connections for both EC and FC. This adds to our understanding of the brain functional alterations in patients with MDD.

## Acknowledgments

This work was supported by the National Natural Science Foundation of China (No. 82372039) and the Natural Science Foundation of Hunan Province of China (No. 2023JJ30695). We thank Yan et al. for sharing and providing the REST-meta-MDD Consortium dataset, and we are grateful for resources from the High-Performance Computing Center of Central South University. Editing assistance was provided by our scientific writer Nancy Linford, Ph.D. (City of Hope National Medical Center).

## Figures and Tables

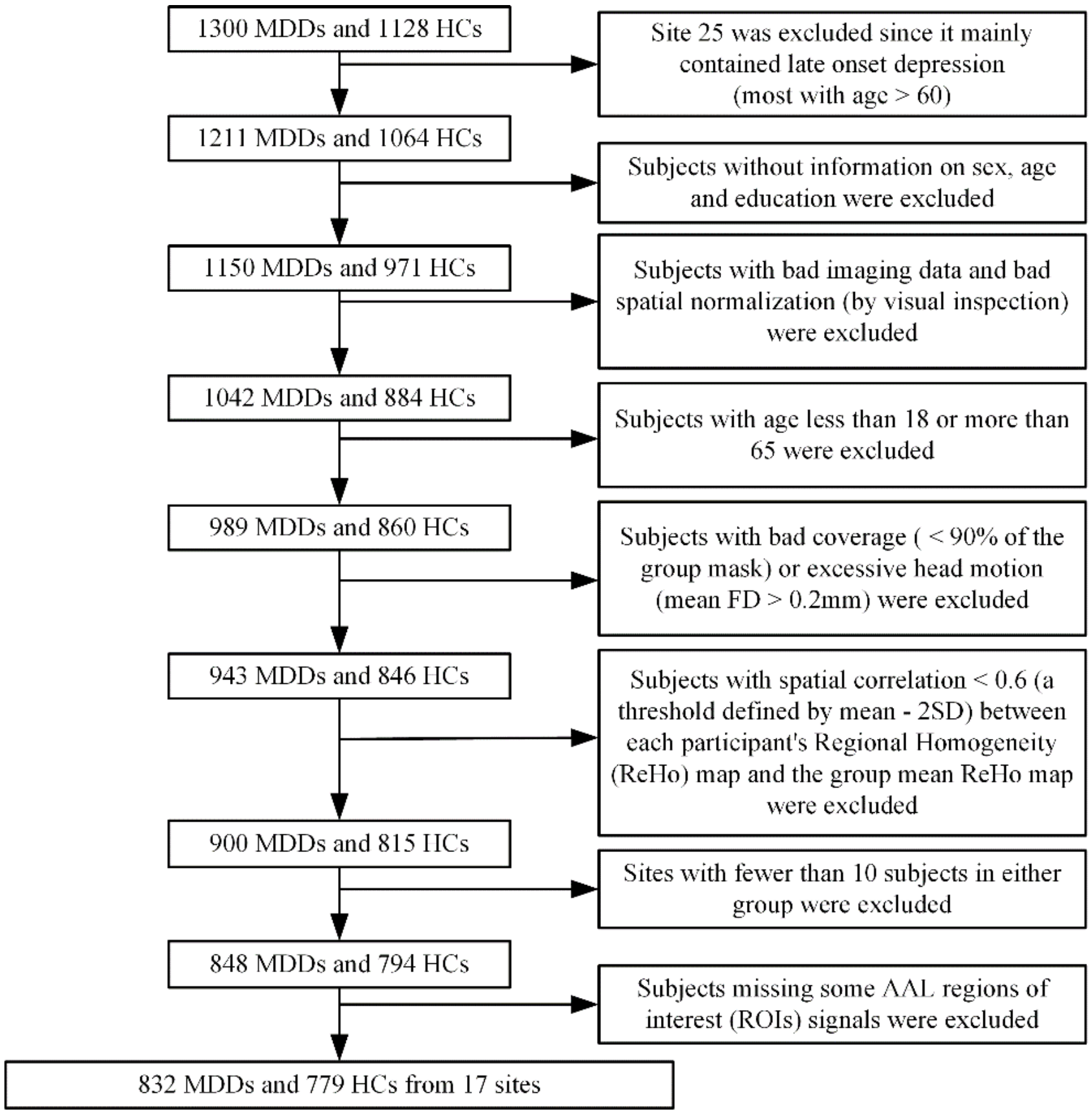

**Figure 1. Flow chart for case screening and exclusion.** AAL, automated anatomical labeling; HC, healthy control; MDD, major depressive disorder; SD, standard deviation; FD, Framewise Displacement.

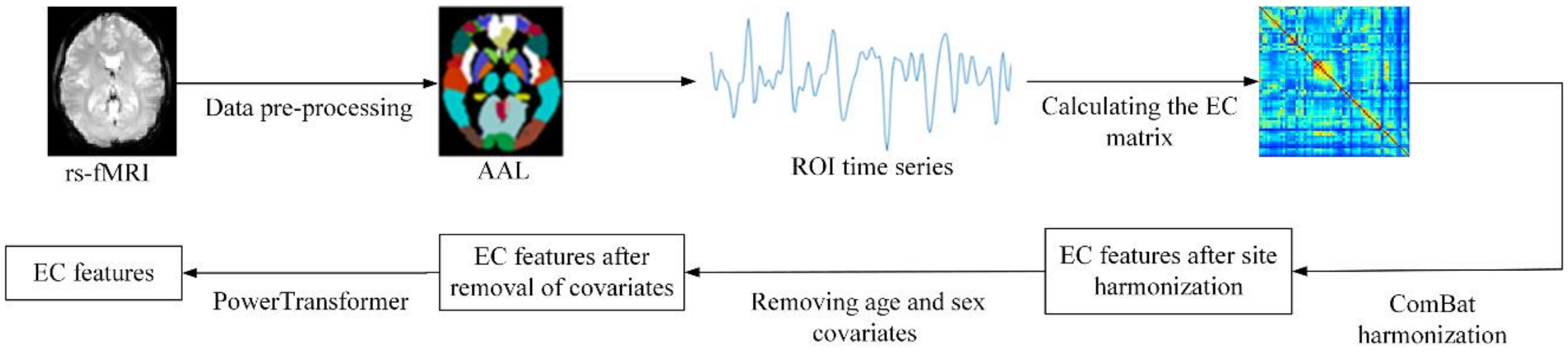

**Figure 2. Flow chart for extracting effective connectivity features.** AAL, automated anatomical labeling; EC, effective connectivity; ROI, region of interest; rs-fMRI, resting state functional magnetic resonance imaging.

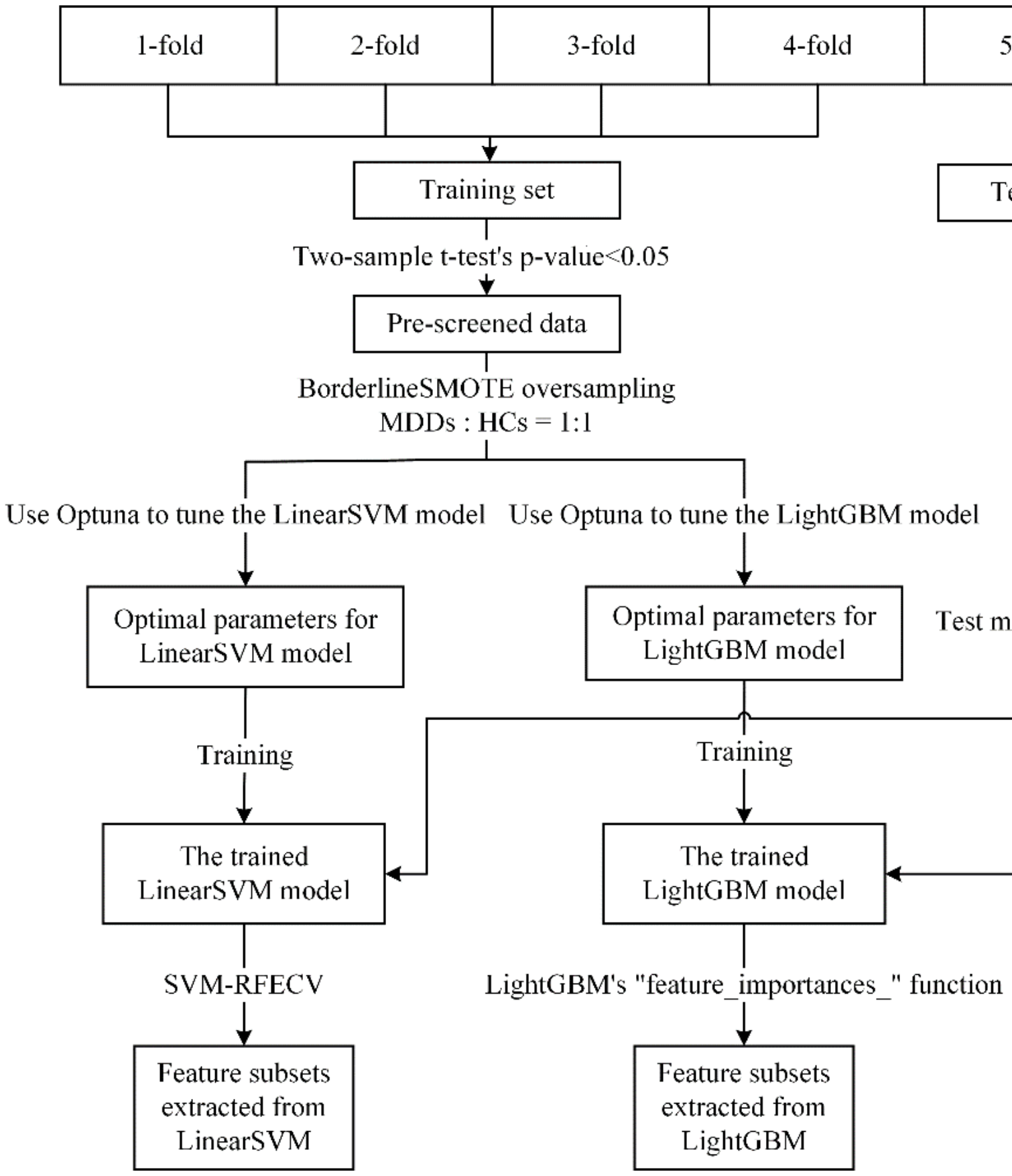

**Figure 3. Flow chart for extraction of important feature subsets.** HC, healthy control; MDD, major depressive disorder.

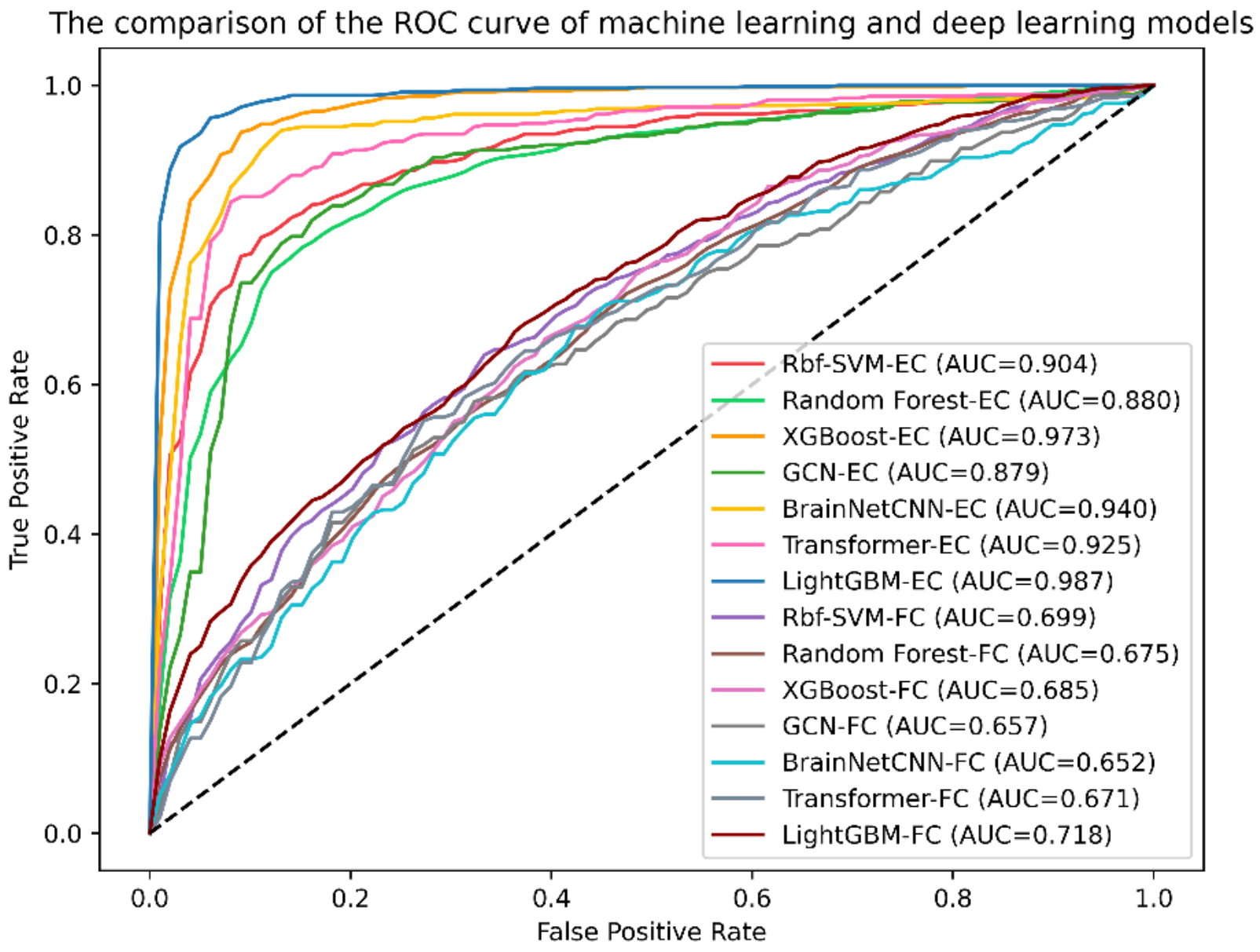

**Figure 4. Comparison of the ROC curves for machine learning and deep learning models.** AUC, area under the ROC curve; ROC, receiver operating characteristics. AUC measures the overall performance of the model by calculating the area under the ROC curve. A higher AUC value, closer to 1, indicates a better model performance. False Positive Rate: ratio of false positive predictions to the total number of actual negative instances, indicating the rate at which the model incorrectly classifies negative instances as positive. True Positive Rate (also known as sensitivity or recall): ratio of true positive predictions to the total number of actual positive instances, indicating the rate at which the model correctly classifies positive instances. Corresponds to Table 1 in the main text.

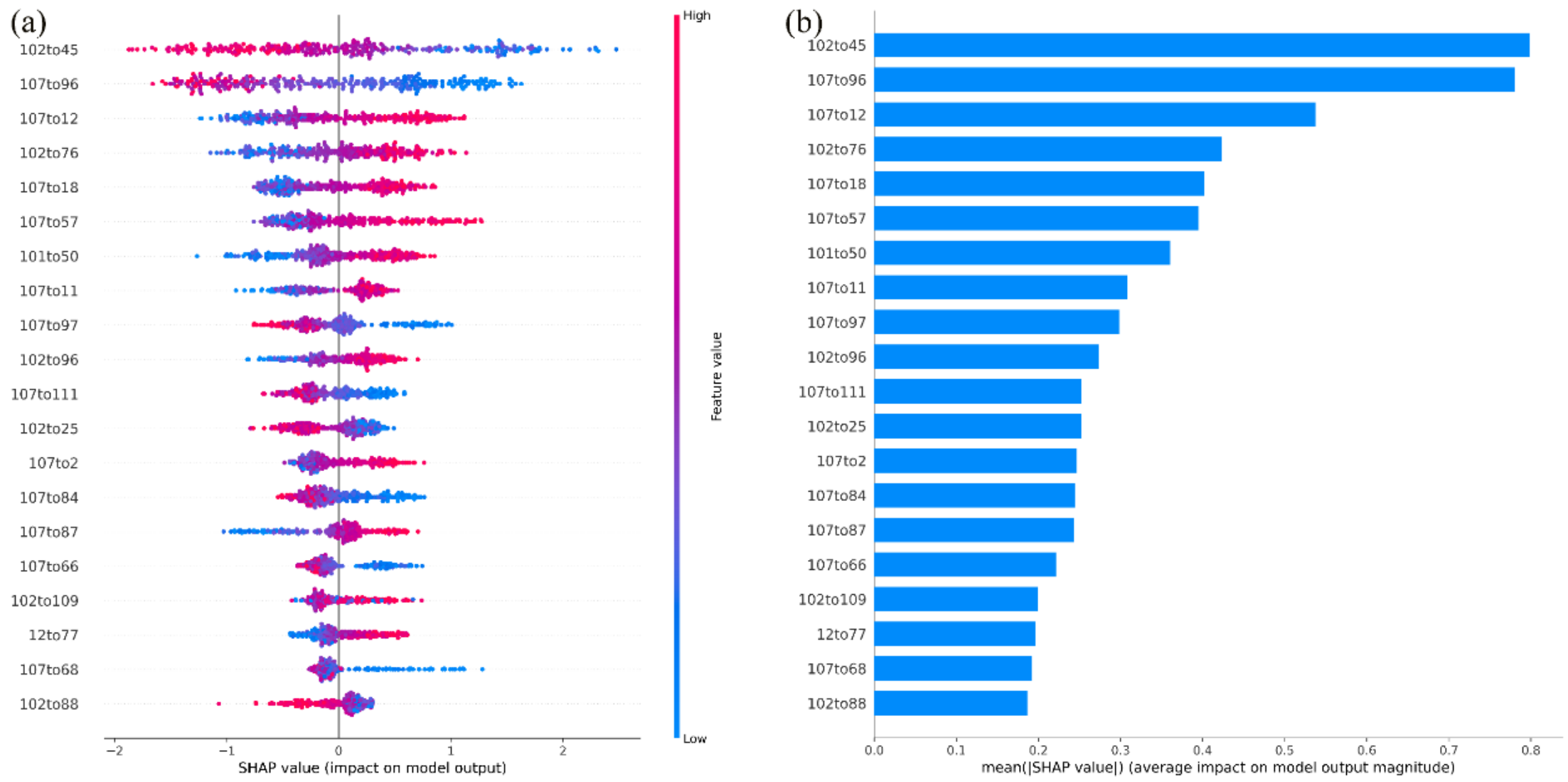

**Figure 5.** **First fold in five-fold result:** **Contribution of important effective connectivity features.** (A) Shown are the Shapley Additive exPlanations (SHAP) values for each of the features, organized by importance from top to bottom. SHAP value < 0 indicates classification as a healthy control (HC), and SHAP value >0 indicates classification as a major depressive disorder (MDD) case. Each point represents a sample's SHAP value, and the point's color depth represents the feature value size. (B) Shown is the mean |SHAP value| of each feature, organized by importance from the top down. A larger mean |SHAP value| indicates a higher contribution of this feature to the model classification, and greater importance. The number of each feature name is the region of interest (ROI) number from the automated anatomical labeling (AAL) atlas. 102to45: EC of Cerebellum_7b_R and Cuneus_L; 107to96: EC of Cerebellum_10_L and Cerebellum_3_R; 107to12: EC of Cerebellum_10_L and Frontal_Inf_Oper_R; 102to76: EC of Cerebellum_7b_R and Pallidum_R; 107to18: EC of Cerebellum_10_L and Rolandic_Oper_R; 107to57: EC of Cerebellum_10_L and Postcentral_L; 101to50: EC of Cerebellum_7b_L and Occipital_Sup_R; 107to11: EC of Cerebellum_10_L and Frontal_Inf_Oper_L; 107to97: EC of Cerebellum_10_L and Cerebellum_4_5_L; 102to96: EC of Cerebellum_7b_R and Cerebellum_3_R; 107to111: EC of Cerebellum_10_L and Vermis_4_5; 102to25: EC of Cerebellum_7b_R and Frontal_Med_Orb_L; 107to2: EC of Cerebellum_10_L and Precentral_R; 107to84: EC of Cerebellum_10_L and Temporal_Pole_Sup_R; 107to87: EC of Cerebellum_10_L and Temporal_Pole_Mid_L; 107to66: EC of Cerebellum_10_L and Angular_R; 102to109: EC of Cerebellum_7b_R and Vermis_1_2; 12to77: EC of Frontal_Inf_Oper_R and Thalamus_L; 107to68: EC of Cerebellum_10_L and Precuneus_R; 102to88: EC of Cerebellum_7b_R and Temporal_Pole_Mid_R.

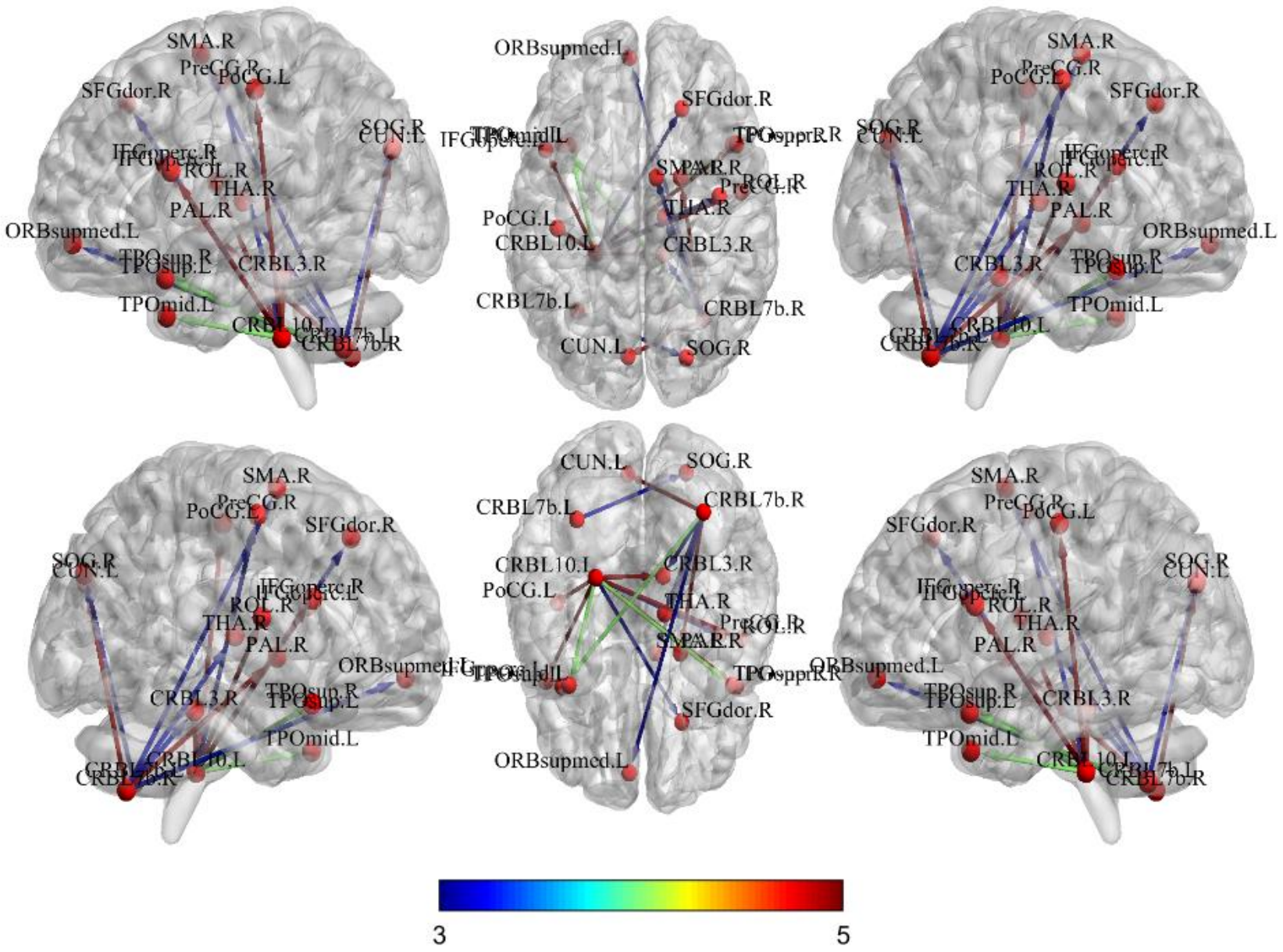

**Figure 6. Brain visualization of high-contribution effective connectivity features.** The color bar indicates the number of occurrences of EC features in the five subsets of high-contribution effective connectivity (EC) features. PreCG.R: Precentral_R; SFGdor.R: Frontal_Sup_R; IFGoperc.L: Frontal_Inf_Oper_L; IFGoperc.R: Frontal_Inf_Oper_R; ROL.R: Rolandic_Oper_R; SMA.R: Supp_Motor_Area_R; ORBsupmed.L: Frontal_Med_Orb_L; CUN.L: Cuneus_L; SOG.R: Occipital_Sup_R; PoCG.L: Postcentral_L; PAL.R: Pallidum_R; THA.R: Thalamus_R; TPOsup.L: Temporal_Pole_Sup_L; TPOsup.R: Temporal_Pole_Sup_R; TPOmid.L: Temporal_Pole_Mid_L; CRBL3.R: Cerebellum_3_R; CRBL7b.L: Cerebellum_7b_L; CRBL7b.R: Cerebellum_7b_R; CRBL10.L: Cerebellum_10_L.

**Table 1. Comparison of the classification performance of machine learning and deep learning models.**

| Model | | Accuracy (%) | Sensitivity (%) | Specificity (%) |
|---|---|---|---|---|
| Machine learning | Rbf-SVM-FC | 64.87% | 66.71% | 62.89% |
| | Rbf-SVM-EC | 83.30% | 84.62% | 81.90% |
| | Random Forest-FC | 62.33% | 62.86% | 61.74% |
| | Random Forest-EC | 89.39% | 88.70% | 90.11% |
| | XGBoost-FC | 62.88% | 64.06% | 61.62% |
| | XGBoost-EC | 91.74% | 91.10% | 92.42% |
| Deep learning | GCN-FC | 60.42% | 66.32% | 57.34% |
| | GCN-EC | 82.26% | 81.84% | 83.16% |
| | BrainNetCNN-FC | 62.15% | 66.81% | 57.18% |
| | BrainNetCNN-EC | 91.18% | 90.36% | 92.05% |
| | Transformer-FC | 62.66% | 64.75% | 60.42% |
| | Transformer-EC | 86.73% | 83.96% | 89.71% |
| Our methods | LightGBM-FC | 65.24% | 66.59% | 63.79% |
| | **LightGBM-EC** | **92.43%** | **91.47%** | **93.45%** |

EC, effective connectivity; FC, functional connectivity.

**Table 2. Classification performance of two classifiers before and after feature selection.**

| Models | Features before feature selection | | | | Features after feature selection | | | |
|---|---|---|---|---|---|---|---|---|
| | Number of features | Accuracy | Sensitivity | Specificity | Number of features | Accuracy | Sensitivity | Specificity |
| LinearSVM | 13340 | 77.15% | 77.76% | 76.50% | 97 | 86.84% | 87.51% | 86.14% |
| LightGBM | 13340 | 92.43% | 91.47% | 93.45% | **97** | **94.35%** | **93.52%** | **95.25%** |

**Notes**: Feature selection was performed using a two-sample t-test and LightGBM's "feature_importances_" function.

**Table 3. Generalizability of highly discriminative effective connectivity features.**

| Models | The REST-meta-MDD Consortium | | | | The DecNef Project Brain Data Repository | | | |
|---|---|---|---|---|---|---|---|---|
| | Number of features | Accuracy | Sensitivity | Specificity | Number of features | Accuracy | Sensitivity | Specificity |
| LinearSVM | 97 | 86.84% | 87.51% | 86.14% | 97 | 83.57% | 85.49% | 82.63% |
| LightGBM | 97 | 94.35% | 93.52% | 95.25% | 97 | 94.74% | 90.59% | 96.75% |

**Notes**: Feature selection was performed using a two-sample t-test and LightGBM's "feature_importances_" function.

**Table 4.** **Literature summary for classification of major depressive disorder using functional connectivity or effective connectivity features (arranged in increasing sample size).**

| Reference | Size of sample | Feature | Classifier | Accuracy | Sensitivity | Specificity |
|---|---|---|---|---|---|---|
| Geng et al. [28] | 24 MDD; 24 HC | FC and EC 272 ROIs (including the AAL atlas and Brainnetome atlas [39]) | Non-LinearSVM, LinearSVM, KNN and LR | 91.67% | 88.00% | 88.00% |
| Sen et al. [6] | 49 MDD; 33 HC | FC between 85 ROIs (the freesurfer cortical parcel-lation atlas [38]); network attribute features extracted from FC | LinearSVM | 79.00% | 86.00% | 70.00% |
| Xu et al.[40] | 98 MDD; 63HC | HTR1A/1B methylation+FC between 116 ROIs (the AAL atlas) | Random Forest | 81.78% | 92.01% | 68.29% |
| Chen et al. [35] | 162 MDD; 38 HC | FC between 142 ROIs (the Dosenbach atlas [36], excluding the cerebellum) | SVM | 88.50% | 80.86% | 92.76% |
| Yamashita et al. [16] | 385 MDD; 849 HC | FC between 379 ROIs (the Glasser's 379 surface-based parcellations [37]) | SVM | 68.00% | 62.00% | 73.00% |
| Qin et al. [31] | 821MDD; 765HC | FC between 160 ROIs (the Dosenbach atlas [36]) | GCN | 81.50% | 83.40% | 80.00% |
| Dai et al. [10] | 832 MDD; 779 HC | FC between 116 ROIs (the AAL atlas, including the cerebellum); network attribute features extracted from FC | LinearSVM, LightGBM | 68.90% | 71.75% | 65.84% |
| Shi et al. [17] | 1021 MDD; 1100 HC | FC between 90 ROIs (the AAL atlas, excluding the cerebellum) | XGBoost | 72.80% | 72.00% | 73.90% |
| Gallo et al.[68] | 1249 MDD; 1249 HC | FC between 112 ROIs(the Harvard-Oxford atlas [69]) | LinearSVM, Rbf-SVM, GCN | 61.27% | 61.38% | 61.31% |
| **Ours** | **832 MDD;** | **EC between 116** | **LightGBM** | **94.35%** | **93.52%** | **95.25%** |

| | | | |
|---|---|---|---|
| | **779 HC** | **ROIs (the AAL atlas, including the cerebellum);** | |

**Notes**: AAL, automated anatomical labeling; EC, effective connectivity; FC, functional connectivity; HC, healthy control; KNN, k-nearest neighbor; LR, logistic regression; MDD, major depressive disorder; ROIs: regions of interest.

**Table 5. High-contribution effective connectivity features.**

| ROI-FROM | ROI-TO | Across the network | MDDs(avg) | HCs(avg) | P-value |
|---|---|---|---|---|---|
| Cerebellum_10_L | Rolandic_Oper_R | Yes | 0.4677 | -0.4995 | <0.0001 |
| Cerebellum_7b_R | Cuneus_L | Yes | -0.0920 | 0.0983 | 0.0001 |
| Cerebellum_10_L | Frontal_Inf_Oper_L | Yes | 0.4713 | -0.5034 | <0.0001 |
| Cerebellum_10_L | Postcentral_L | Yes | 0.3286 | -0.3509 | <0.0001 |
| Cerebellum_10_L | Cerebellum_3_R | Yes | -0.4256 | 0.4546 | <0.0001 |
| Cerebellum_7b_R | Pallidum_R | Yes | 0.1064 | -0.1136 | <0.0001 |
| Cerebellum_10_L | Frontal_Inf_Oper_R | Yes | 0.4785 | -0.5110 | <0.0001 |
| Cerebellum_7b_R | Temporal_Pole_Sup_L | Yes | -0.1540 | 0.1645 | <0.0001 |
| Cerebellum_10_L | Temporal_Pole_Mid_L | Yes | 0.2418 | -0.2582 | <0.0001 |
| Cerebellum_10_L | Temporal_Pole_Sup_R | Yes | -0.3159 | 0.3374 | <0.0001 |
| Cerebellum_7b_R | Supp_Motor_Area_R | Yes | 0.0909 | -0.0971 | 0.0002 |
| Cerebellum_10_L | Frontal_Sup_R | Yes | -0.4122 | 0.4403 | <0.0001 |
| Cerebellum_7b_R | Thalamus_R | Yes | 0.1923 | -0.2054 | <0.0001 |
| Cerebellum_10_L | Precentral_R | Yes | 0.1501 | -0.1603 | <0.0001 |
| Cerebellum_7b_R | Frontal_Med_Orb_L | Yes | -0.0980 | 0.1046 | <0.0001 |
| Cerebellum_7b_L | Occipital_Sup_R | Yes | 0.0934 | -0.0998 | 0.0001 |
| Cerebellum_7b_R | Cerebellum_3_R | Yes | 0.1134 | -0.1211 | <0.0001 |

**Notes:** "Across the network" refers to whether the two ROIs involved in the feature belong to the same resting-state network. MDDs (avg) and HCs (avg) refer to the average of the feature values (The feature values have been processed by age & sex covariate regression and ComBat.) of MDDs and HCs in the REST-meta-MDD Consortium dataset, respectively. P-value is the result of a two-sample t-test for feature values between MDDs and HCs in the REST-meta-MDD Consortium dataset.